\documentclass{IEEEtran}
\usepackage{cite}
\usepackage{amsmath,amssymb,amsfonts}
\usepackage{subfigure}
\usepackage{algorithmic}
\usepackage{graphicx}

\usepackage{textcomp}
\usepackage{xcolor}
\usepackage{booktabs}
\usepackage{authblk}
\usepackage{multirow}
\usepackage{textcomp}

\usepackage{cite}
\usepackage{amsmath,amssymb,amsfonts}
\usepackage{algorithmic}
\usepackage{graphicx}
\usepackage{textcomp}

\def\BibTeX{{\rm B\kern-.05em{\sc i\kern-.025em b}\kern-.08em
    T\kern-.1667em\lower.7ex\hbox{E}\kern-.125emX}}
\begin{document}
\title{Software-Defined Edge Computing: A New Architecture Paradigm to Support IoT Data Analysis}
\author{Di Wu$^1$, Xiaofeng Xie$^1$, Xiang Nie$^1$, Bin Fu$^1$, Hanhui Deng$^1$, Haibo Zeng$^2$, and Zhijin Qin$^3$
\\
$^1$ Key Laboratory for Embedded and Network Computing of Hunan Province, Hunan University, China
\\$^2$ Department of Electrical and Computer Engineering, Virginia Tech, USA
\\$^3$ EECS, Queen Mary University of London, UK
}

\maketitle

\begin{abstract}
The rapid deployment of Internet of Things (IoT) applications leads to massive data that need to be processed. These IoT applications have specific communication requirements on latency and bandwidth, and present new features on their generated data such as time-dependency.  Therefore, it is desirable to reshape the current IoT architectures by exploring their inherent nature of communication and computing to support smart IoT data process and analysis. We introduce in this paper features of IoT data, trends of IoT network architectures, some problems in IoT data analysis, and their solutions. Specifically, we view that software-defined edge computing is a promising architecture to support the unique needs of IoT data analysis. We further present an experiment on data anomaly detection in this architecture, and the comparison between two  architectures for ECG diagnosis. Results show that our method is effective and feasible.
\end{abstract}


\section{Introduction}
\label{sec:introduction}
Internet of Things (IoT) applications in smart cities have generated massive data. According to a report from Cisco, the number of connected IoT mobile devices (MDs) will reach 11.6 billion by 2021, which will generate 49 exabytes mobile data traffic per month \cite{b1}. These IoT devices connect with location-dependent heterogeneous access points (APs) deployed on the edge of networks. The heterogeneity of APs leads to the coexistence of various communication technologies (e.g., WiFi, Bluetooth, Cellular). It is very challenging for network administrators to manage such a large scale of IoT devices associated with heterogeneous networks. Meanwhile, the massive data generated from IoT applications create unexpected problems which can not be directly addressed through traditional approaches designed for low-rate  systems.

To meet the need for low-latency or even timing guarantees in emerging IoT applications, current IoT architecture must be redesigned from the physical layer to the application layer in a bottoms-up fashion to fully meet these challenging requirements. In addition, ubiquitous IoT devices often are leveraged as an infrastructure to collect massive spatial-temporal data, hence big data analysis is becoming an important aspect of IoT application. First, quality of the collected data is critical for the accuracy of data analysis results. Second, as IoT data are generated by distributed smart devices and sensors, time and spatial dependencies are the most unique feature compared to traditional data. Therefore, novel methods for analyzing and processing IoT data are developed and deployed on IoT,
which also introduces new requirements to IoT architecture.

To tame the above challenging issues, smart IoT architectures are in need to improve the capability of IoT and adapt to the IoT data analysis. Specifically, with the emerging of new network technologies, Software Defined Network (SDN) \cite{b2} and Edge Computing \cite{b3} provide promising paradigms to reshape IoT network architectures. SDN helps to manage edge computing resources, in return, edge computing can enable SDN to manage IoT in a distributed manner\cite{b4}. Within the smart architecture enabled by the integration of SDN, edge and IoT components, agile data analysis methods may be supported to fully explore the features of IoT data.

\section{IoT: State of The Art, Trends, Features and Issues}
\subsection{State of The Art}
IoT is different from traditional networks, which generally consists of perception layer, network layer and application layer. Massive perceptual infrastructures are deployed to collect data, and then the collected data are transmitted to remote cloud servers to be stored and processed. Usually, sensor nodes\cite{b5} or smart devices transmit the data to the remote cloud through IoT gateways or access points. The application layer in the cloud analyzes the data and utilizes the analysis results. Based on this common IoT architecture, previous work aimed at extending novel network technologies to meet the demands from its special scenarios.


For example, Galluccio et al. \cite{b6} presented an application-oriented solution to make sensor nodes programmable so that developers can customize the sensor platform according to application requirements. Furthermore, to realize smart urban sensing, Liu et al. \cite{b7} proposed a software-defined IoT architecture which decouples sensing applications and the physical infrastructure. It is a flexible solution to support various application requirements and convenient management of the physical infrastructure. Besides, managing geographically distributed, heterogeneous networking infrastructures is a key technical challenge in a dynamic environment. Qin et al., \cite{b8} proposed a software-defined approach to dynamically achieve differentiated quality levels to different IoT tasks in heterogeneous wireless networking scenarios. As for access control and mobility management, Wu et al. \cite{b9} presented solutions for ubiquitous flow control and mobility management by maintaining a controller network based on structured overlay of hash ring.

\subsection{Trends in IoT Architectures}
\textbf{Software-defined IoT:} Due to the heterogeneous nature of the IoT, SDN can be an ideal solution to overcome the limitations of current IoT network infrastructures. The key idea is to separate the control plane from the data plane, which improves flexibility, openness, and programmability and hence the usability of the network. OpenFlow \cite{b2} is one of the most important Southbound protocols that implements the SDN concept.

SDN brings many new features into IoT network architectures, such as network programmability, centralized control, and flexible management. First, SDN can simplify the management of IoT architectures. The SDN controller enables IoT devices to be managed centrally through remote configuration services. Second, SDN is inherently extensible and can quickly add new IoT devices. The abstract features of SDN enable IoT applications to access data, analyze and control devices, and add new sensors and network control equipments at the same time, which does not need to expose the underlying infrastructure details. Moreover, SDN supports the dynamic reconfiguration of network equipments and data traffic\cite{b10}, which facilitates the IoT to flexibly adjust components according to changes of data flows.

\textbf{IoT-EDGE-SDN:} Many IoT applications have strict requirements on timing, reliability, and security. Due to constraints of limited bandwidth and computation resources, IoT needs to preload the traffic. Therefore, it needs to deploy a platform, which integrates the capabilities of connectivity, computing, storage, and applications on the edge of networks. The paradigm of edge computing emerges to meet such requirements. In  IoT applications with real-time requirements, such as autonomous driving, control delay must be less than a few milliseconds. If control servers are placed in the cloud, it will not be able to meet the tight control delay requirement. Therefore, certain analysis and control functions need to be placed on the edge of networks to meet the low latency demand.

Integrating IoT with SDN and Edge Computing is promising to achieve complementary benefits. Through the IoT-EDGE-SDN ecosystem, IoT has the potential to make full use of their advantages on ubiquitous sensing and data collection, which could accelerate the growth of IoT applications. In addition to low latency, edge computing can provide distributed management for IoT. Thus, the stability of local system services in IoT can be guaranteed by the cooperation of  distributed intelligent and autonomous systems on the edge. For example, we can push SDN to the network edge and construct a distributed control plane to manage the mobile access of IoT devices\cite{b11}. Furthermore, as substantial computing resources are deployed on edge servers in the form of virtual machines (VMs) and are utilized by IoT, the SDN controllers have the capability to support management of these edge resources.

\subsection{Features of IoT Data}
\begin{figure}[htp]
	\begin{center}
		\begin{tabular}{cc}
			\includegraphics[width=4.2cm]{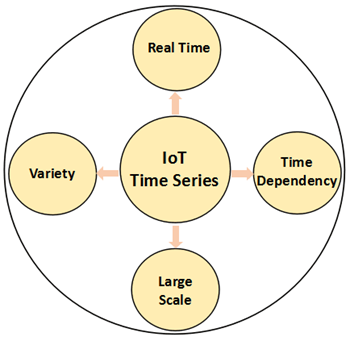} &
			\includegraphics[width=4.2cm]{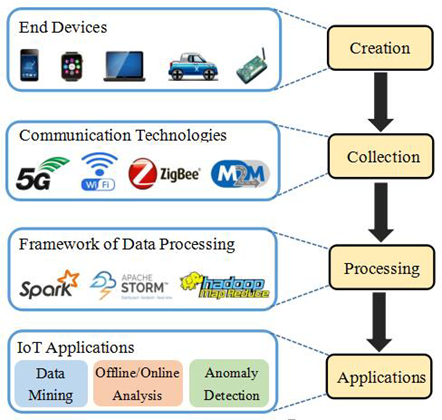}\\
			{\scriptsize (a)} &{\scriptsize (b)}
		\end{tabular}
		\caption{IoT data: (a) Features; (b) Data processing procedure.}
		\label{fig2}
	\end{center}
\end{figure}

Generally in distributed IoT architectures, the IoT devices, such as smart phones and sensors, are deployed in a distributed fashion to collect different types of data. As in Figure~\ref{fig2}(a), data gathered by IoT have the following features: \textbf{(a) Large Scale}. Massive sensor data are gathered by distributed equipments all the time. \textbf{(b) Variety}. There is a large variety of data acquisition devices so that the type of collected data is also diversed in the IoT. \textbf{(c) Time Dependency}. Sensor data in IoT are collected by devices at specific locations and are labeled with time stamps, as the data streams are measurements of continuous physical phenomena. Time dependency within data streams is inherent and thus is one of the most critical features. \textbf{(d) Real Time}. The collected data should be processed in a timely fashion to ensure the validity of the data. Also, if the sensor device is running abnormally, it needs to be detected within a certain amount of time to avoid affecting the normal operation of other devices.

Therefore, IoT data not only helps to obtain the operational status of IoT devices, but also provides abundant opportunities for other IoT applications. Figure~\ref{fig2}(b) shows the IoT data processing procedure, including data creation, collection, processing and specific applications. Diverse sensors or smart devices collect data at the edge of the network, and the IoT data are transmitted to remote servers via 5G, Wi-Fi, ZigBee, etc. The data are stored and processed in the server using various data processing frameworks. The processed results could be later used by specific applications, e.g., data mining, offline/online analysis, anomaly detection, for smart IoT management and services.

\subsection{Issues in IoT}
Though previous works have presented important contributions to develop new communication and computing paradigms for the extension of existing IoT design, there are still many challenges to be addressed. In this section, we focus on the discussion of relevant issues from the perspective of designing an agile IoT architecture for data processing and analysis.

\textbf{IoT Integrating Software-defined Networking and Edge Computing:} To better manage IoT and provide low latency services, new computing paradigms including SDN and edge computing need to be merged into  IoT. However, the design of hybrid architecture integrating IoT with SDN and Edge is challenging. The deployment of network components should be considered to make the IoT system scalable and robust. The cloud center and edge equipments need to coordinate to meet different service requests from IoT devices.

\textbf{Data Collection and Sensing:} For smart IoT management and decision, it is essential to obtain valuable information from massive spatial-temporal data generated by IoT devices. Data redundancy and duplication may impact the accuracy of data processing and analysis operations, hence  obtaining high-quality data is crucial. In an IoT environment, we have to also consider the cost such as energy and computation load. Therefore, how to improve the efficiency of data collection is a challenge.

\textbf{Data Processing:} In smart IoT systems, massive sensor data are gathered by distributed equipments. With a large amount of IoT data, how to find an efficient and reliable scheme to process these data has become a critical issue. Usually specific requirements of IoT applications in different deployment scenario are different as well. However, most IoT applications have specific expectation on the time delay during data processing, since many collected IoT data should be processed immediately to ensure the validity of these data.

\textbf{Data Analytics:} Collected IoT data contains noisy information. If data mining is applied directly to the original data, it cannot necessarily ensure accuracy and reliability of data analysis. Therefore, it is challenging to effectively preprocess the IoT data under the condition that the key information of the IoT data is not lost. Meanwhile, IoT data are time-dependent, so the data mining procedure should consider the recursive relationship among data events. Since the time interval between some events is relatively large, we have to solve a long-term time-dependent problem.

\section{Potential Solutions}
\subsection{An IoT-EDGE-SDN Architecture for IoT Data Analysis}
\begin{figure}[htp]
	\centerline{\includegraphics[width=\columnwidth]{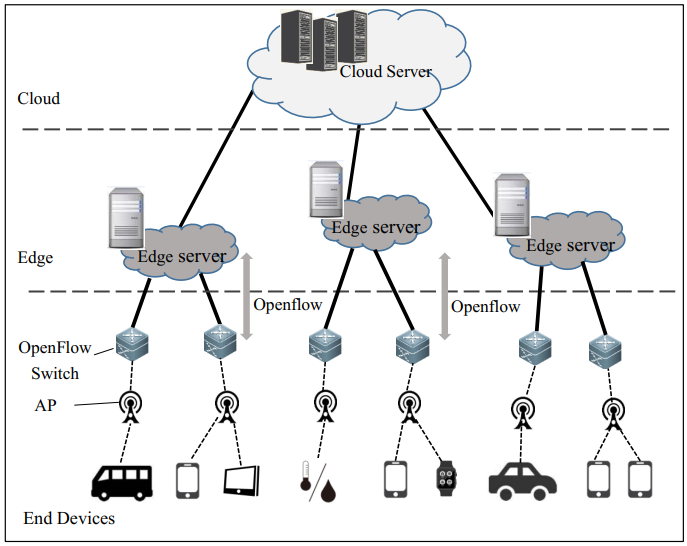}}
	\caption{A three-tier IoT-EDGE-SDN architecture.}
	\label{fig3}
\end{figure}

As discussed in previous sections, distributed architecture is the development trend of IoT. To better manage IoT devices, collect data information, process and analyze IoT data, and provide low latency services, the current IoT architecture needs to be redesigned. Inspired by new computing paradigms SDN and Edge Computing, Figure~\ref{fig3} conceptually illustrates a three-tier IoT architecture, which integrates SDN and Edge Computing to support distributed controllers and connected switches built on the OpenFlow framework. The IoT-EDGE-SDN architecture consists of the IoT device layer, the edge layer and the cloud layer.

\textbf{IoT Device Layer:} Different smart devices and sensors are associated with heterogeneous access points, which are connected to OpenFlow switches to request various types of services. The OpenFlow switches are connected with local SDN controllers. The IoT devices can interact with SDN controllers through the southbound protocol supported by OpenFlow switches. For heterogeneous network devices and access methods, the southbound interface protocols need to be extended to support smarter and more friendly multi-network access. Generally, IoT devices can collect data which are then transmitted to edge servers to be stored and processed.

\textbf{Edge Layer:} The SDN controllers are deployed closely to the network edge, which divides the network into several partitions. The controllers can centrally manage the IoT devices within their partitions. The resources of storage, computing, and communication are also available in the edge server rather than all in the cloud. The computing resource typically lands on the edge server in the form of a virtual machine (VM) or container. The controller can also support the unified management functions of various VMs, including image file management, resource management, and life cycle management. Besides, the edge server can support third-party applications or platforms to perform specific data analysis and processing tasks on the IoT data. For example, we can deploy the TensorFlow platform, which significantly facilitates the realization of machine learning on IoT data analysis. The processing results are consumed by applications and may also be transmitted to the cloud for offline batch processing.

\textbf{Cloud Layer:} Although the management of IoT in the cloud server is weakened, it still has powerful computing resources and centralized control on the distributed edge servers including SDN controllers. The cloud server can provide offline analysis and support non-real-time applications. Besides, the cloud server sends decisions and configurations to the edge servers, and handles unsolved tasks uploaded by them.	

\subsection{Data Collection and Sensing}
The data generated by IoT devices needs to be transmitted to edge servers. Some sensing or communication techniques can be used for collecting data, such as wireless sensor network (WSN). The effectiveness of data collection is a key factor in IoT systems, which affects the quality of collected data significantly. Currently, efficient data collection  have been well investigated in sensory systems. There are many ways to improve the data collection efficiency in IoT systems. In general, smart IoT devices are usually shared by different IoT components. Therefore, concurrent data collection is expected. Cheng et al. \cite{b12} presented a delay-aware network structure specifically designed for concurrent data collection  in IoT systems. Li et al.\cite{b13} proposed a method to optimize multimedia data collections in mobile sensing vehicles. This method targets at balancing multimedia data collections, improving the delivery ratio of the multimedia data, and reducing the delay ratio in IoT networks.

Obtaining high quality data can help better conduct data mining and data analysis. As for data sensing, the IoT data that have the spatio-temporal correlation enables advanced sensing techniques. For example, compressive sensing can reduce the sampling rate and the network traffic loads between IoT devices and analysis server. Besides, mobile crowd sensing (MCS) is also an important sensing paradigm for large-scale IoT, which collects the sensing information from many participating mobile devices.  Han et al. \cite{b14} proposed a hybrid framework to compensate for inadequate sensing opportunities solely provided by incentive mechanisms, which combines mobile devices with static sensor nodes to generate uniformly distributed space-time data under the constraint of field coverage.

\subsection{Processing in IoT Data}
Due to the explosive growth of IoT data, the traditional high-performance services no longer meet the daily processing needs. Furthermore, the diversity of IoT data leads to different response times, where some data require real-time processing, while others require further data mining. Distributed parallel computing frameworks have been used to efficiently process data. For example, MapReduce \cite{b15}, which is a computational framework for big data parallel processing, is used to make offline batch analysis and calculation of IoT data. MapReduce can automatically allocate and execute tasks, and then collect computational results on cluster nodes.

As for iterative operations such as data mining or machine learning in IoT data applications, we can use the Spark \cite{b15} parallel computing framework, which is an open source cluster computing environment similar to Hadoop. The difference is that, during the calculation of Spark, the intermediate output of jobs can be stored in memory, so it is no longer necessary to read from and write to Hadoop distributed file system. Similarly, Spark can provide quick and interactive queries, because its resilient distributed datasets are also stored in memory.

Some IoT data in specific scenarios are required to be processed in real time to achieve millisecond-level responses, such as automatic pilot. For these scenarios, Storm \cite{b16} is a suitable real-time big data processing framework. It guarantees that every message will be processed at least once. In a small cluster, it can handle millions of messages per second.

\subsection{Analysis in IoT Data}
IoT data have very complex features. To ensure the accuracy and reliability of results, it is necessary to preprocess the data. If the sampling frequency is very high, the down sampling operation needs to be carried out first. When the number of features of the collected data is too large, it is necessary to reduce the dimension of features. The commonly used dimension reduction operations include principal component analysis (PCA), Linear Discriminant Analysis, Lasso, etc. Among them, PCA is a linear projection that maps high-dimensional data to low dimensional space representation, and expects the largest variance on the projected dimension of data. To speed up the convergence of the PCA algorithm, min-max normalization can be used to normalize the IoT data. Moreover, missing values may exist in the data, which requires some padding operations, for example, the use of the mean, mode and median filling, or exploring the relationship between the features, or establishing a random forest prediction model for filling.

\begin{figure}[htp]
	\centerline{\includegraphics[width=\columnwidth]{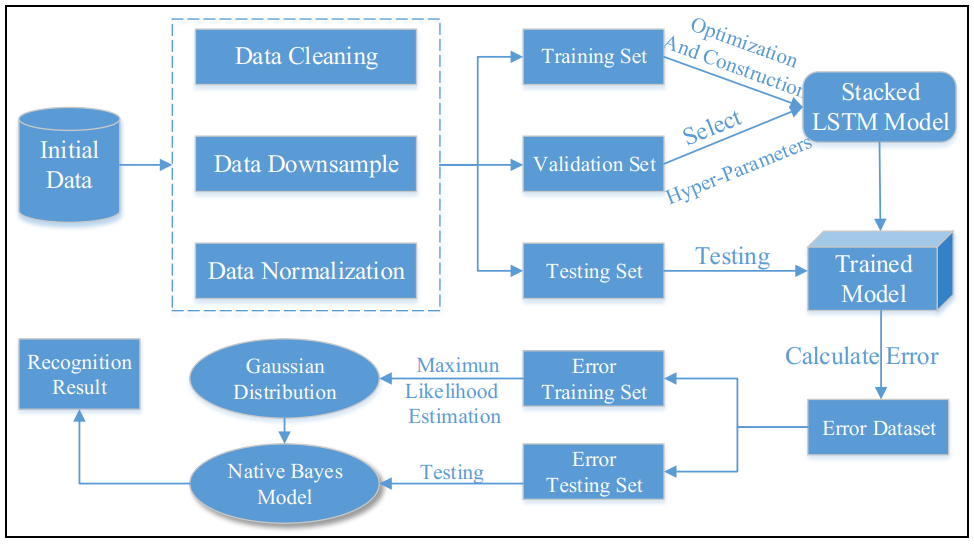}}
	\caption{Anomaly detection in IoT data.}
	\label{fig4}
\end{figure}

For the analysis of time-dependent features in IoT data, a typical method is the Autoregressive Moving Average Model (ARIMA) \cite{b16}. The ARIMA model transforms non-stationary time series into stationary time series, and then predicts future values from the past and present values. Furthermore, there are some sequence models, such as conditional random field model, Kalman filter and Markov model, dealing with sequential data, but unable to learn long-range dependencies. The long short-term memory (LSTM) neural network \cite{b17} is a variant of the recurrent neural network (RNN), which can effectively solve the problem of long-term time-dependency by introducing a set of memory units. LSTM networks have been widely used in many sequence learning tasks\cite{b18}, such as speech recognition and machine translation. Due to the excellent predictive performance of LSTM on time series data, we propose an LGNB (LSTM with Gaussian and Naive Bayes) model to detect anomaly of IoT data. As illustrated in Figure~\ref{fig4}, we use the training set to build the stacked LSTM model. Then, we import the test set into the trained predictive model to construct the error data set. Furthermore, the error training set is used to build Naive Bayes model \cite{b19} of Gaussian distribution to identify the abnormal behavior.

\section{Evaluation}
In order to validate the effectiveness of our approach to analyzing IoT data,  we perform two sets of experiments. The first is anomaly detection on IoT time series data, and the second is a comparative study on two different architectures for electrocardiogram (ECG) diagnosis. One of the structures is the IoT-EDGE-SDN  and the other one uses cloud platform only.

\subsection{Anomaly Detection on IoT Time Series Data}

We use Google's deep learning platform, Tensorflow, to implement our algorithm. The edge computing platform configured with Nvidia GTX1070 GPUs is used to accelerate our training on the model. To validate our model, we compare LGNB model with the LSTM NN model and multi-layer perception (MLP) model, respectively. The LSTM NN model includes an input layer, two hidden layers with LSTM memory blocks and a classification layer. The MLP model has an input layer, multiple hidden layers, and a classification layer. Here, we construct a perceptron model with three hidden layers, which have 10, 20, 10 units, respectively. The cost function is the cross entropy loss function for both models. A real-world time series data set, power data set, is selected to perform the experiment. All models are trained on 80 percent of the data and tested on 10 percent of the data, and the remaining 10 percent is used for validation. We use the adaptive gradient algorithm (Adagrad) to train each model 1000 epoch. To prevent over-fitting, we use regularization techniques and dropout to reduce the complexity of these models. In addition, we use 5-fold cross validation to select the hyper-parameters of these models.

\begin{figure}[htp]
	\begin{center}
		\begin{tabular}{cc}
			\includegraphics[width=4.2cm]{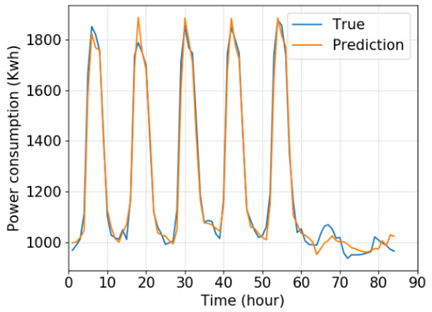} &
			\includegraphics[width=4.2cm]{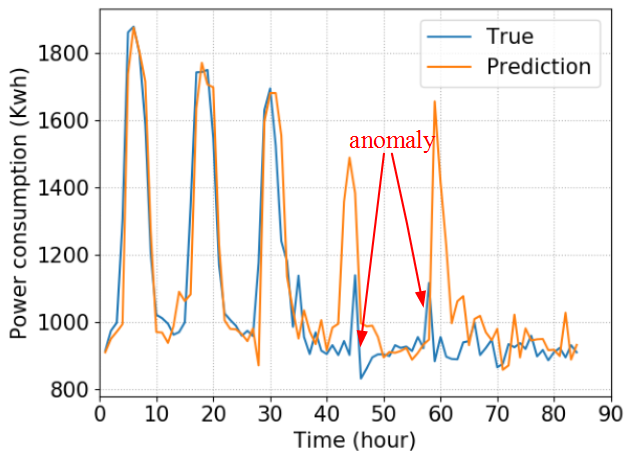}\\
			{\scriptsize (a)} &{\scriptsize (b)}
		\end{tabular}
		\begin{center}
			\includegraphics[width=\columnwidth]{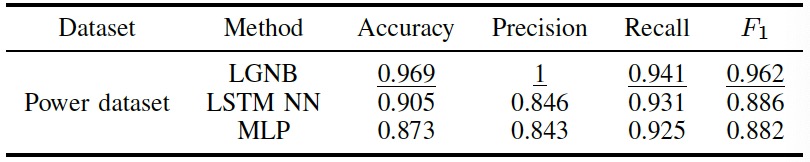}
			{\scriptsize (c)}
		\end{center}
		\caption{The predicted results of the model in the power data set: (a) normal sample; (b) abnormal sample; (c) The performance of each model under four classification indicators (Accuracy, Precision, Recall, $F_1)$.}
		\label{fig5}
	\end{center}
\end{figure}


Our data set is from a user's power data for a period of one year. It was collected every 15 minutes each day. We down-sample the original power data for each week, and the resulting data constitutes the input samples for our model. Under the normal circumstances, power consumption will be relatively high in the weekdays, and relatively low on the weekend. From Figures~\ref{fig5}(a) and~\ref{fig5}(b), the trend of power consumption shows 5 peaks in the weekdays, and a trough appears in the weekend. If a sample had troughs in the weekdays, or wave crest appeared in the weekend, we could consider it as an anomaly. In addition, the data are noisy, so the peak does not appear exactly at the same time of each day.

In order to evaluate our LGNB model, we use four metrics: accuracy, precision, recall, and $F_1$ to compare the models. Generally speaking, accuracy is enough to judge a model whose goal is to classify. Since our goal is to identify whether the sample is abnormal, precision and recall are important metrics to evaluate our model. Precision rate is mainly used to judge whether the classifier can correctly identify the anomaly, i.e., it mainly focuses on identifying abnormal samples. The recall rate is the proportion of abnormal samples being identified. $F_{\beta}$ is a combination of the previous two metrics. If $\beta$ value is less than 1, it indicates the recall rate is more important. On the contrary, the precision rate has a greater impact on the model quality assessment. This experiment uses $F_1$, i.e., precision rate and recall rate are equally important.


Figure~\ref{fig5}(c) shows the performance of our models under four classification metrics. We use the underscore to mark the highest value for each metric. For the power data set, we find that the autoregressive correlation coefficient is more than 0.5 when the delay is 10. Therefore, we know that this is a data set with a long-term dependency on the time, and its current data are inextricably linked to the data in front of it. As in Figure~\ref{fig5}(c), the model with the hidden layer of the LSTM unit is generally better than the general hidden layer in the results of each of metrics. However, the data set has a certain periodicity (with a period of one week), feature extraction is also easy to be implemented for the ordinary neural network model. The performance of the MLP model is also good whose recall rate is 92.5\%.

\subsection{Comparative Study on ECG Diagnosis}

\begin{figure}[htp]
	\begin{center}
		\begin{tabular}{c}
			\includegraphics[width=\columnwidth]{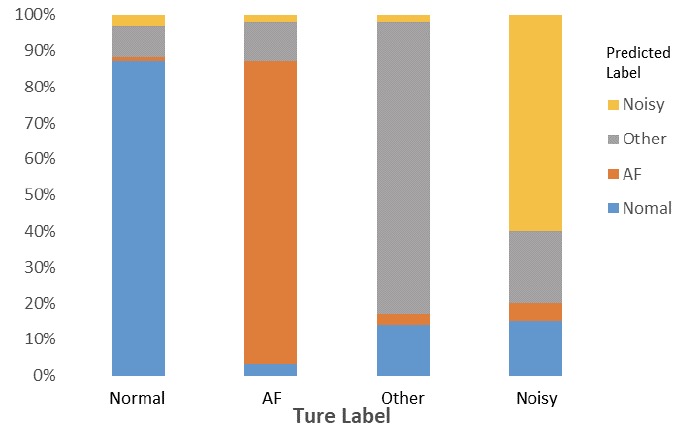} \\
			{\scriptsize (a)}
		\end{tabular}
		\begin{center}
			\includegraphics[width=\columnwidth]{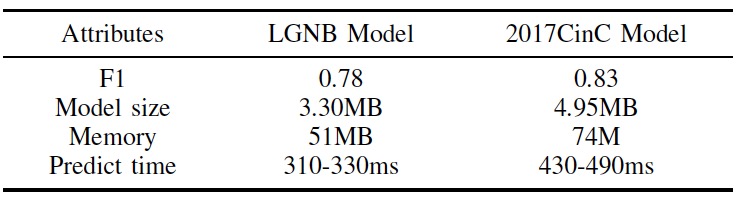}
			{\scriptsize (b)}
		\end{center}
		\caption{The experiments results of two different architectures for ECG diagnosis: (a) The prediction accuracy of each category; (b) The differences between the two models in precision, required storage at runtime, memory, and prediction time.}
		\label{fig6}
	\end{center}
\end{figure}

Here, we assess the performance of two ECG diagnosis models  on the edge device, which is  a smart mobile phone with snapdragon 660 processor. As shown in Figure~\ref{fig6}(a), we test the prediction accuracy of the model in four categories, including normal rhythm (N), AF rhythm (A), other rhythm (O), and noisy recordings ($\sim$)).  In the figure, ``Normal'' outperforms the other three categories. The  second highest accuracy rate is achieved by ``AF''  84\%. Moreover, the other three categories of ECG can be easily misjudged as ``Other''. The ratio of ``AF'' to ``Other'' is 11\%, which is quite high. ``Noisy'' achieves the lowest prediction accuracy, and most of them are classified into ``Other'' and ``Normal''.

Afterwards, the accuracy of the model in terms of predicting $F_1$, model size,  memory, and prediction time. Here, $F_1$ measures of the accuracy of the model in predicting atrial fibrillation. The comparison results between LGNB and the model in 2017CinC~\cite{b20} are presented in Figure~\ref{fig6}(b). We can see that our model requires 1.65MB less than the 2017CinC model in terms of the model size even though the achieved F1 is slightly lower. While for the memory at runtime, our LGNB models is 23M less than the 2017CinC model. Moreover, the required predication time for our LGNB model is significantly less the 2017CinC model. Therefore, we could conclude that our model outperforms the benchmark with smaller size of the storage space occupied by the model, less memory consumption, and shorter inferred time at the cost of slight degradation on F1.

In the following experiments, we focus on performance comparison of the two structures in terms of delay. In the experiment, the  time length from the data generation to the feedback of diagnostic results to the user is measured under different request volumes, which includes the data transfer time and model prediction time.

\begin{figure}[t!]
\centerline{\includegraphics[width=3.5in]{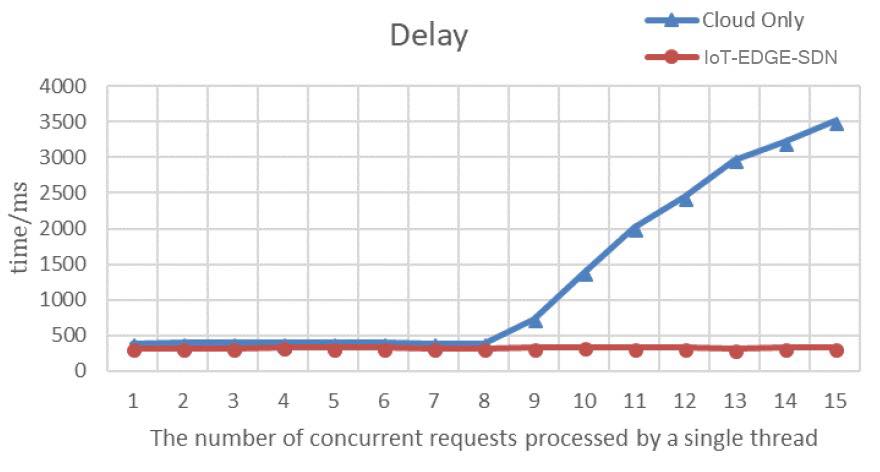}}
\caption{Comparison of delay between the two  architectures.}
\label{fig}
\end{figure}

As shown in Fig.~\ref{fig}, with  the increasing request volume of each thread, the cloud structure faces bottlenecks in performance and network I/O, resulting in a gradual increase in latency. The delay reaches approximately 393 ms when the number of concurrent requests for each thread of the cloud platform is less than 7. When it is further increased from 7 to 8, we can achieve the minimum delay, 384.7 ms. In this case, the could platform achieves the highest processing efficiency as the request rate and the highest rate that can be handled by a single thread of the cloud platform are close, which avoids the frequency activation of the thread. Once the number of requests exceeds the processing capability of a single thread, the request will be queued. However, for most cases, health monitoring data are  generated continuously and usually should be processed in real time. Once a request is queued, a large amount of data will be in congestion. As a result, the  delay is significantly increased and then the real-time diagnostic capability is lost.

Different from the cloud, edge devices are distributed and deployed locally, such as at user's home. Edge devices usually have certain data computing capabilities. With the increasing number of users, similar as the cloud platform, the delay of our IoT-EDGE-SDN model can be managed, which is stabilized at about 320 ms. Usually, we can have multiple edge devices at home, which gives us the confidence that our IoT-EDGE-SDN model is an efficient and reliable solution for healthcare data processing.

\section{Conclusion}
The blooming of Internet of things generates large scale of time-series data in need of effective and efficient analysis. In order to cope with the great traffic pressure for the current IoT architecture, SDN and Edge Computing are promising computing paradigms to support the increasingly complex IoT architecture. This article has proposed a new three-tier IoT architecture and discussed some issues in architecture design, data collection and sensing, data processing, and data analysis. Our experimental evaluation confirms that our methods to analyze and process IoT data provide promising results.

\end{document}